\documentclass[aps,prb,twocolumn,showpacs,superscriptaddress,groupedaddress]{revtex4-1}

\usepackage{amsmath}
\usepackage{amssymb}   
\usepackage{graphicx}
\usepackage{multirow}  
\usepackage{braket}
\usepackage{CJK}

\begin{document}

\title{Topological invariants in terms of Green's function for the interacting Kitaev chain}

\author{Zhidan Li$^1$}
\author{Qiang Han$^{1,2}$}
\address{$^1$Department of Physics, Renmin University of China, Beijing 100872, China}
\address{$^2$Beijing Key Laboratory of Opto-electronic Functional Materials and Micro-nano Devices, Renmin University of China, Beijing 100872}

\date{\today}

\begin{abstract}
The one dimensional closed interacting Kitaev chain and the dimerized version are studied.
The topological invariants in terms of Green's function are calculated by the density matrix renormalization group method and the exact diagonalization method.
For the interacting Kitaev chain, we point out that the calculation of topological invariant in the charge density wave phase must consider the dimerized configuration of the ground states. The variation of topological invariant are attributed to the poles of eigenvalues of the zero-frequency Green's functions.
For the interacting dimerized Kitaev chain, we show that the topological invariant defined by the Green's functions can distinguish more topological nonequivalent phases than the fermion parity. \\


\end{abstract}


\maketitle

The Kitaev chain model, as a prototype model of 1D topological superconductor, has attracted enormous interest since it was proposed in Kitaev's seminal work.~\cite{Ref_Kitaev_0} The most intriguing property of the Kitaev chain is the appearance of the Majorana zero modes (MZM) in a chain with open boundary condition. The robustness of gapless boundary modes is protected by the nontrivial topology of the bulk band structure. This intimate connection between the bulk topological properties and the gapless boundary modes  is known as bulk-boundary correspondence.~\cite{Ref_classification_NON} Hence the topological $Z_2$ invariant, which characterizes the topology of the Kitaev chain, can be calculated either from the single-particle Bogoliubov-de Gennes Hamiltonian~\cite{Ref_Kitaev_0, Ref_Alicea_0} in momentum space or from the emergence of the boundary modes.

However, when interactions are introduced, the bulk topological invariant can no longer be built on the basis of the single particle picture.
It is an interesting issue how to define and compute the topological invariant for the interacting Kitaev chain. To date, most theoretical works investigated the effect of interaction on the topology properties by examining the variation of the zero-bias Majorana peaks as well as the fermion parity and degeneracy of the ground state for the open interacting Kitaev chain.~\cite{Ref_interaction_0,Ref_interaction_1,Ref_interaction_2,Ref_interaction_3,Ref_interaction_4,Ref_interaction_5,Ref_interaction_6,Ref_interaction_7,Ref_interaction_8} There is still a lack of the direct calculation of the bulk topological invariant of the closed interacting Kitaev chain. To characterize the topologies of the many-body ground states beyond the single particle description, people proposed~\cite{Ref_Volovik_2003, Ref_SCZhang_2010, Ref_SCZhang_2012a, Ref_SCZhang_2012b, Ref_SCZhang_2012c, Ref_Wang_2013, Ref_Gurarie_2011, Ref_Gurarie_2012} the Green's function approach to topological insulators and superconductors. The topological invariants are expressed in terms of zero-frequency Green's function~\cite{Ref_SCZhang_2012a, Ref_SCZhang_2012b, Ref_Gurarie_2012} which replaces the role of single particle Hamiltonian and is a useful tool to explore the topological properties of the interacting Kitaev chain.

In this paper, firstly, we calculate the bulk topological invariant of the interacting Kitaev chain numerically and find that the Green's function formalism works well in the topological superconducting (TSC) phase. However, the formula must be generalized for the charge density wave (CDW) phase where the translational symmetry is broken.
Secondly, we show that for both the noninteracting and interacting Kitaev chain, the variation of the topological invariant can be attributed to the poles of eigenvalues of the zero-frequency Green's function.~\cite{Ref_Volovik_2003, Ref_Gurarie_2011, Ref_Gurarie_2012, Ref_Yoshida_2014}
Finally, we reveal that the Green's function formalism is powerful in calculating the topological invariant of an interacting dimerized Kitaev chain.

The Hamiltonian of the closed interacting Kitaev chain with periodic boundary condition is written as,
\begin{align}\label{Eq_Ham_Kitaev-U}
     H   =  & \sum_{j=1}^{L} (-t c_{j}^{\dagger}c_{j+1} - \Delta c_{j}^{\dagger}c_{j+1}^{\dagger} + h.c.)
              - \mu \sum_{j=1}^{L} (c_{j}^{\dagger}c_{j} - \frac{1}{2}), \nonumber\\
            & +  U \sum_{j=1}^{L} (n_{j}-\frac{1}{2})(n_{j+1}-\frac{1}{2}),
\end{align}
where $c_{j}^{\dagger}$ denotes the creation operator on the $j$-th site.
$t$, $\Delta$, $\mu$ and $U$ are the nearest-neighbor hopping integral, the $p$-wave pairing potential, the chemical potential and the nearest-neighboring interaction, respectively.
When $U=0$, the model is reduced to the original Kitaev chain, which has a nontrivial topological number when $\left|\mu\right|<2t$.~\cite{Ref_Kitaev_0, Ref_Alicea_0}
In this paper $t$ is chosen as the unit of energy and $\Delta$ is set as real and positive without loss of generality. We focus on the half filling case with $\mu=0$.

\begin{figure}[htbp]
\begin{center}
  \includegraphics[width=7.5cm, angle=0]{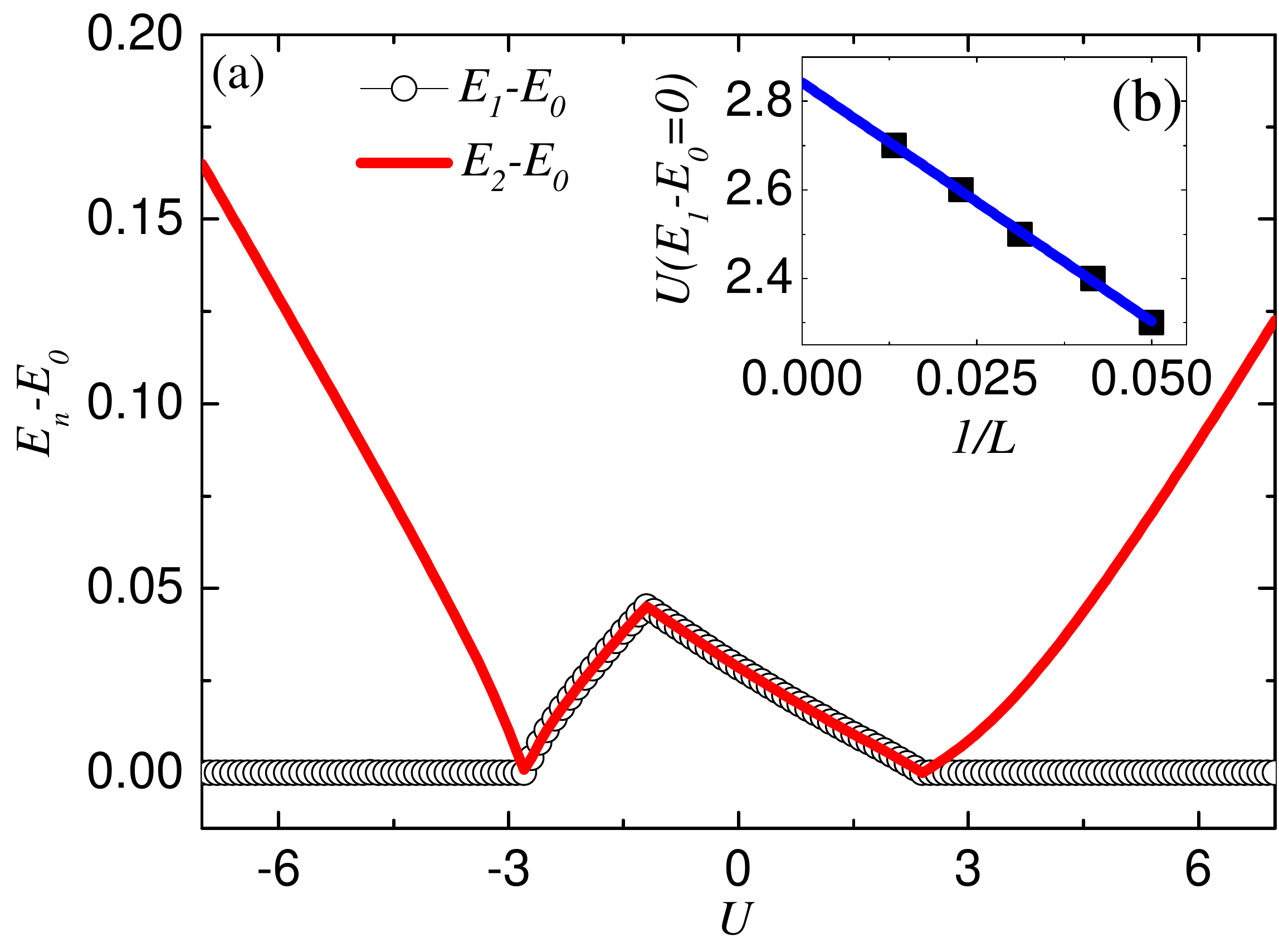}
\end{center}
 \caption{(Color online) Many-body energy spectrum of the closed interacting Kitaev chain as a function of the interaction $U$.
          The black circle and red line denote the first ($E_1$) and second ($E_2$) excitation energies relative to the ground state energy respectively.
          The inset is the finite-size scaling of the right boundary between $E_1-E_0\neq0$ and $U(E_1-E_0=0)$.
          the parameters of model are chosen as $\Delta=0.4$, $L=20 \thicksim 100$.
         }
 \label{Fig_En}
\end{figure}

The phase diagram of the interacting Kitaev chain at half filling is quite simple.~\cite{Ref_DP_of_XYZ_0,Ref_DP_of_XYZ_1}
There are three phases: the schr\"{o}dinger-cat (CAT) phase~\cite{Ref_interaction_8} for $U<-2(t+\Delta)$, the CDW phase for  $U>2(t+\Delta)$,
and the TSC phase in between.~\cite{Ref_DP_of_XYZ_0}
Fig.~\ref{Fig_En} shows the many-body energy spectrum of a closed chain as a functions of the interaction $U$ when $\Delta=0.4$ obtained by density matrix renormalization group (DMRG) method.
The ground states are two-fold degenerate in the CAT and CDW phases even for a finite-size closed chain.
The ground state of the TSC phase is non-degenerate (no MZMs for a closed chain).
It can be seen from Fig.~\ref{Fig_En} that the phase transition point between the CAT (TSC) phase and the TSC (CDW) phase is $U_c\approx-2.8~(+2.5)$ for $L=28$.
The inset of Fig~\ref{Fig_En} indicates that the phase boundary between the TSC and CDW phases is $U=2.8$ in the thermodynamic limit.
Therefore the two phase boundaries obtained by DMRG are in agreement with the exact values, $U_c=\pm 2(t+\Delta) = \pm 2.8$ for $\Delta=0.4$.

The interacting Kitaev chain has the chiral symmetry, whose topological invariant can be calculated by the zero-frequency Nambu Green's function,~\cite{Ref_SCZhang_2012b, Ref_SCZhang_2012c, Ref_Gurarie_2011, Ref_Gurarie_2012}
\begin{equation}\label{Eq_N1_0}
    N_1 =  \text{tr} \int_{-\pi}^{\pi} \frac{dk}{4\pi i} ~\hat{\mathcal{S}}~g^{-1}(k)~\partial_k g(k),
\end{equation}
where $\hat{\mathcal{S}}$ denotes the chiral operator, tr tracing for the indices of Nambu spinors,
\begin{equation}\label{Eq_Nambu_1}
  \Psi=(c_k,c_{-k}^{\dagger})^{T}.
\end{equation}
The corresponding zero-frequency Nambu Green's function is written as,~\cite{Ref_SCZhang_2012c, Ref_Gurarie_2012}
\begin{equation}\label{Eq_G_N_0}
  g(k) =
          \langle \langle
            \Psi | \Psi^{\dagger}
          \rangle \rangle_{0}
        =
            \left(
             \begin{array}{cc}
               G_{c_k,c_k^{\dagger}}(0)              &  G_{c_k,c_{-k}}(0)  \\
               G_{c_{-k}^{\dagger},c_k^{\dagger}}(0) &  G_{c_{-k}^{\dagger},c_{-k}}(0)
             \end{array}
           \right),
\end{equation}
where,
\begin{align}\label{}
    &\langle \langle \hat{A} | \hat{B} \rangle \rangle_{z} = G_{\hat{A},\hat{B}}(z) \nonumber\\
    &=\langle \psi_0 | \hat{A} \frac{1}{z+E_{0}-\hat{H}} \hat{B} | \psi_0 \rangle
                                  + \langle \psi_0 | \hat{B} \frac{1}{z-E_{0}+\hat{H}} \hat{A} | \psi_0 \rangle,
\end{align}
where $\hat{A}$ and $\hat{B}$ are placeholders for operators, $| \psi_0 \rangle$ the ground states, $z\in\mathbb{C}$.
The notation $\langle \langle \hat{A} | \hat{B} \rangle \rangle_{z}$ was firstly introduced by Bogoliubov.~\cite{Ref_Bogoliubov_0}

Due to the properties of the Green's functions, the following equations can be proven easily:
\begin{align}\label{Eq_G_N_1}
  g(k) &= \left(
                         \begin{array}{cc}
                             g_{11}^{x}   &  ig_{12}^{y}  \\
                             -ig_{12}^{y} &  -g_{11}^{x}
                         \end{array}
                     \right),
\end{align}
where $g_{ij}^{x}$ ($g_{ij}^{y}$) is the real (imaginary) part of the matrix element of $g(k)$.
In the Nambu representation defined by Eq.~\eqref{Eq_Nambu_1}, $\hat{\mathcal{S}} = \sigma_x$, where $\sigma_i$ ($i=x,y,z$) is the Pauli matrix.
Introducing a unitary transformation,
\begin{align}\label{trans}
  U & = \frac{1}{\sqrt{2}}
        \left(
         \begin{array}{cccc}
          1  & 1  \\
          -i & i  \\
         \end{array}
        \right),
\end{align}
it can be obtained that $\hat{\mathcal{S}}=U^\dagger\sigma_z U$ and
\begin{equation}\label{gtrans}
    g(k)
      = U^\dagger \left(
         \begin{array}{cccc}
          0          & Z(k)  \\
          Z^*(k) & 0     \\
         \end{array}
        \right) U
\end{equation}
From Eq.~\eqref{Eq_G_N_1}, we have $Z(k)=Z_x(k)+iZ_y(k)=g_{12}^{y}+ig_{11}^{x}$. Here $Z_{x(y)}(k)$ is the real (imaginary) part of $Z(k)$. Then after substituting Eq.~(\ref{gtrans}) into Eq.~(\ref{Eq_N1_0}), the topological invariant can be written as,
\begin{align}\label{Eq_N1_1}
    N_1 = & - \frac{1}{2\pi i} \int_{-\pi}^{\pi} dk ~Z^{-1}(k)\partial_kZ(k)
        =   - \frac{1}{2\pi i} \oint_{C} \frac{dZ} {Z},
\end{align}
where $C$ denotes the closed contour circled by $Z(k)$ as $k$ varying from $-\pi$ to $\pi$.
From the Eq.~\eqref{Eq_N1_1}, it can be found that the topological invariant $N_1$ can be interpreted as the winding number of $Z(C)$ around the origin on the complex plane.
Numerically, $g_{11}^{x}(k)$ and $g_{12}^{y}(k)$ and $Z(k)$ at each $k$ can be calculated,
and then the topological invariant $N_1$ can be obtained from the locus of $Z(k)$.~\cite{Ref_Yoshida_2014}

Furthermore, people pointed out that the change of topological number $N_1$ can be attributed to the poles or zeros of eigenvalues of zero-frequency Green's functions,~\cite{Ref_Volovik_2003, Ref_Gurarie_2011, Ref_Gurarie_2012, Ref_Yoshida_2014, Ref_Slager_2015} as argued in the following. From Eqs.~(\ref{trans}) and (\ref{gtrans}), we have the
relation,
\begin{equation}\label{Eq_det_gk}
    \prod_n \lambda_n(k) = \text{det} ~g(k) = -[Z_x^2(k) + Z_y^2(k)],
\end{equation}
where $\lambda_n(k)$ is the $n$-th eigenvalue of $g(k)$.
From Eqs.~(\ref{Eq_N1_1}), one can see that with the varying of system parameters the topological invariant changes and therefore the topological phase transition happens on the condition that certain $Z(k)$ moves to infinity or becomes zero. Finally the behavior of $Z(k)$ at the phase transition is connected to the infinite or vanishing eigenvalues of $g(k)$ according to Eq.~(\ref{Eq_det_gk}).

We take the noninteracting Kitaev chain as an example to illustrate that the change of topological invariant of a noninteracting model only originates from the poles of the zero-frequency Green's functions.
The momentum-space Hamiltonian of noninteracting Kitaev chain in the Nambu representation defined by Eq.~\eqref{Eq_Nambu_1} is,
\begin{equation}\label{}
    H_0(k)
       =\left(
         \begin{array}{cc}
          - 2t\text{cos}k - \mu    &  -2i\Delta \text{sin}k  \\
          2i\Delta \text{sin}k  &  2t\text{cos}k + \mu      \\
         \end{array}
        \right).
\end{equation}
And $Z(k)$ in Eq.~\eqref{Eq_N1_1} for this noninteracting Kitaev chain is $ Z(k)=[2\Delta \text{sin}k - i (2t\text{cos}k+\mu) ]/\epsilon^2(k)$,
where $\epsilon (k) = \sqrt{(2t\text{cos}k+\mu)^2 + (2\Delta \text{sin}k)^2}$. $\pm Z(k)$ are just the eigenvalues of $g(k)$$=$$-H^{-1}_0(k)$.
Therefore, the change of topological invariant $N_1$ at $\mu=\pm2t$ is due to the poles of $g(k)$ for the noninteracting Kitaev chain.

\begin{figure}[htp]
\begin{center}
  \includegraphics[width=7.5cm, angle=0]{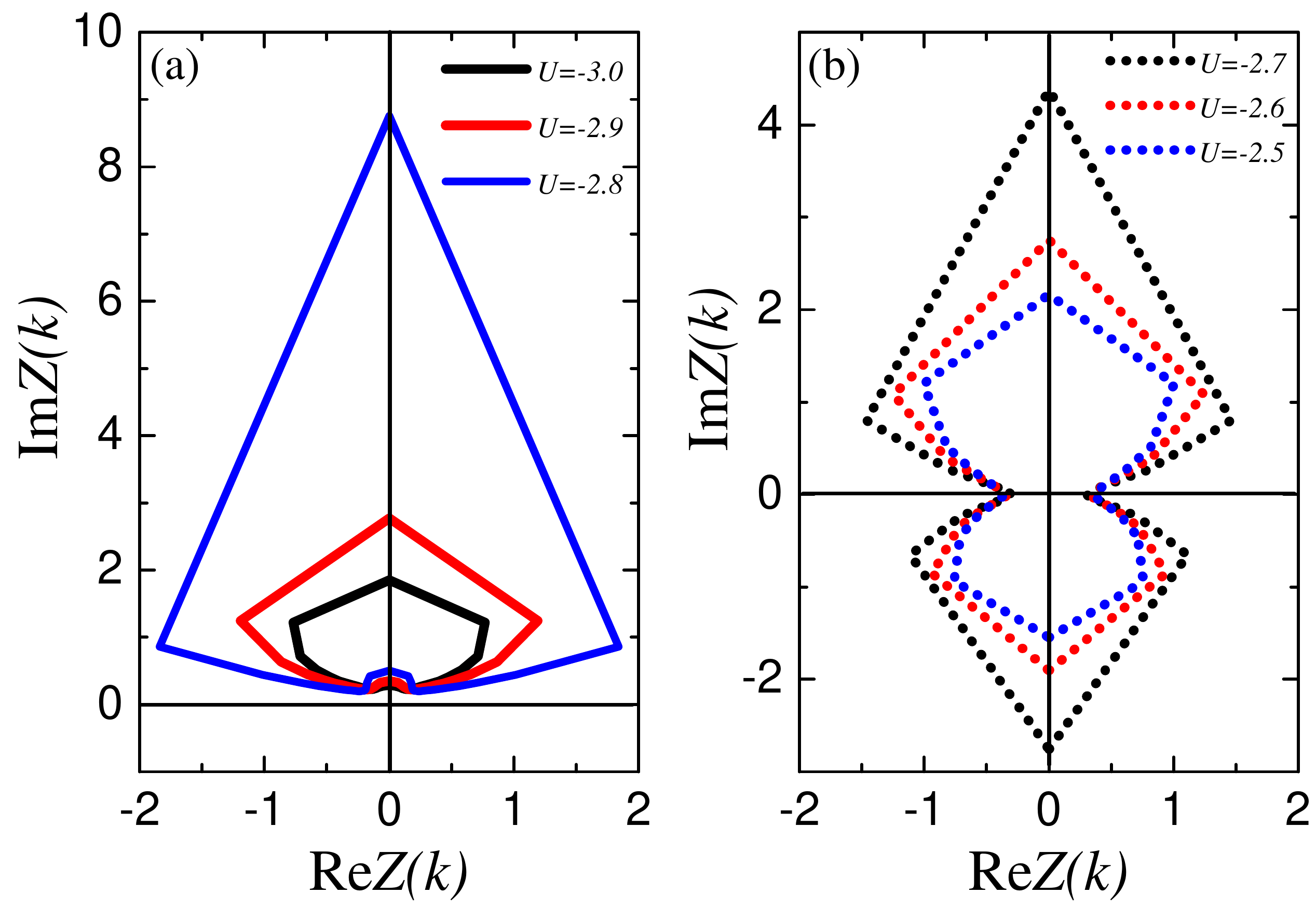}
\end{center}
 \caption{(Color online) The variation of topological invariant from the CAT phase to the TSC phase for the closed interacting Kitaev chain.
          The horizontal (vertical) axis denotes the real (imaginary) part of $Z(k)$ of Eq.~\eqref{Eq_N1_1}.
          The parameters of model are chosen as $\Delta$=0.4, $L$=40.
         }
 \label{Fig_TN_1}
\end{figure}

Next we investigate the topological invariant of the interacting Kitaev chain using the DMRG method. The locus of $Z(k)$ for different $U$ around $-2.8$ is calculated, as shown Fig.~\ref{Fig_TN_1}. Contrary to the trivial CAT phase, $Z(k)$ of the TSC phase winds once counterclockwise around the origin of the complex plane, which indicates $N_1=1$. Therefore the TSC (CAT) phase is topologically nontrivial (trivial). Furthermore, Fig.~\ref{Fig_TN_1}(a) shows that when $U$ varies from $-3.0$ to the phase boundary $U_c\approx-2.8$, $Z(k=\pi)$ increases rapidly towards infinity along the positive direction of the imaginary axis. In comparison, when $U$ varies from $-2.5$ to $U_c$, $Z(k=\pi)$ approaches infinity along the negative direction of the imaginary axis as illustrated in Fig.~\ref{Fig_TN_1}(b). Therefore, there is a pole of eigenvalues of the zero-frequency Nambu Green's functions at $U_c\approx-2.8$, which results in the variation of the topological invariant between the CAT phase ($N_1=0$) and the TSC phase ($N_1=1$).

However, if the above formalism, which is applicable to the TSC and CAT phases, is used to calculate the topological invariant of the CDW phase, an incorrect topological number is obtained with $N_1=-1$ (not shown in the figure) and therefore a generalization for the CDW phase is needed. We note that when $U$ becomes more and more repulsive, the dimerized configurations have more and more weights in the ground state and the system enters into the CDW phase.
Therefore, it is natural to include these dimerized configurations in the calculation of topological invariant of the CDW phase.
We introduce the dimerized Nambu spinor,
\begin{equation}\label{Eq_Nambu_2}
    \Psi^{\prime}=(c_{k,A},c_{k,B},c_{-k,A}^{\dagger},c_{-k,B}^{\dagger})^{T},
\end{equation}
where $A$ and $B$ denote the two sublattice sites in one unit cell. The corresponding dimerized Nambu Green's function at zero frequency is written as,
\begin{align}\label{Eq_G_N_2}
  &g(k)
    =
        \langle \langle
          \Psi^{\prime} | \Psi^{\prime \dagger}
        \rangle \rangle_{0} \nonumber\\
    &=
     \left(
       \begin{array}{cccc}
         g_{11}^x                & g_{12}^x+ig_{12}^y   & ig_{13}^y               & g_{14}^x+ig_{14}^y      \\
         g_{12}^x-ig_{12}^y  & g_{22}^x                 & -g_{14}^x+ig_{14}^y & ig_{24}^y                  \\
         -ig_{13}^y              & -g_{14}^x-ig_{14}^y  & -g_{11}^x               & -g_{12}^x-ig_{12}^y     \\
         g_{14}^x-ig_{14}^y  & -ig_{24}^y               & -g_{12}^x+ig_{12}^y & -g_{22}^x                   \\
       \end{array}
      \right),
\end{align}
where $g_{ij}^x$ $(g_{ij}^y)$ is the real (imaginary) part of the matrix element of $g(k)$. In the Nambu representation defined by Eq.~\eqref{Eq_Nambu_2}, the chiral operator $\hat{\mathcal{S}}=\sigma_x \otimes \tau_0$,
where $\tau_0$ is a $2\times2$ identity matrix. Introducing a unitary transformation $U$,
\begin{align}\label{}
  U & = \frac{1}{\sqrt{2}}
        \left(
         \begin{array}{cccc}
          1  & 1  \\
          -i & i  \\
         \end{array}
        \right)
        \otimes
        \tau_0,
\end{align}
it can be obtained that $\hat{\mathcal{S}} = U^\dagger (\sigma_z\otimes\tau_0) U$, and
\begin{equation}\label{Eq_Qk}
    g(k)
      = U^\dagger \left(
         \begin{array}{cc}
          0         & Q(k)  \\
          Q^\dagger(k) & 0  \\
         \end{array}
        \right) U,
\end{equation}
where $Q(k)$ is a $2\times2$ matrix.
Defining $Z(k) \equiv \text{det} Q(k) = Z_x(k)+i Z_y(k)$, $Z_x(k)$ and $Z_y(k)$ $\in \mathbf{R}$ can be expressed in terms of the Green's functions with the help of Eq.~\eqref{Eq_G_N_2} and~\eqref{Eq_Qk},
\begin{align}\label{}
Z_x(k) &=  {g_{12}^{x}}^2 \!+\! {g_{12}^{y}}^2 \!-\! {g_{14}^{x}}^2 \!-\! {g_{14}^{y}}^2 \!-\! g_{11}^{x}g_{22}^{x} \!+\! g_{13}^{y}g_{24}^{y} ,  \nonumber \\
Z_y(k) &=  2g_{14}^{x}g_{12}^{y} - 2g_{12}^{x}g_{14}^{y} + g_{11}^{x}g_{24}^{y} + g_{22}^{x}g_{13}^{y}.
\end{align}
Finally, the topological invariant can be calculated by,
\begin{align}\label{Eq_N1_2}
    N_1 = & \text{tr} \int_{-\pi}^{\pi} \frac{dk}{4\pi i} ~\hat{\mathcal{S}}~ g^{-1}(k)~\partial_k g(k)
        =  - \frac{1}{2\pi i} \oint_{C} \frac{dZ}{Z}.
\end{align}
Likewise, $g_{11}^{x}$, $g_{12}^{x}$, $g_{12}^{y}$, $g_{13}^{y}$, $g_{14}^{x}$, $g_{14}^{y}$, $g_{22}^{x}$, $g_{24}^{y}$ and $Z(k)$
for every $k$ can be calculated by DMRG,
and then the topological invariant $N_1$ are obtained from the locus of $Z(k)$.

\begin{figure}[htp]
\begin{center}
      \includegraphics[width=7.5cm, angle=0]{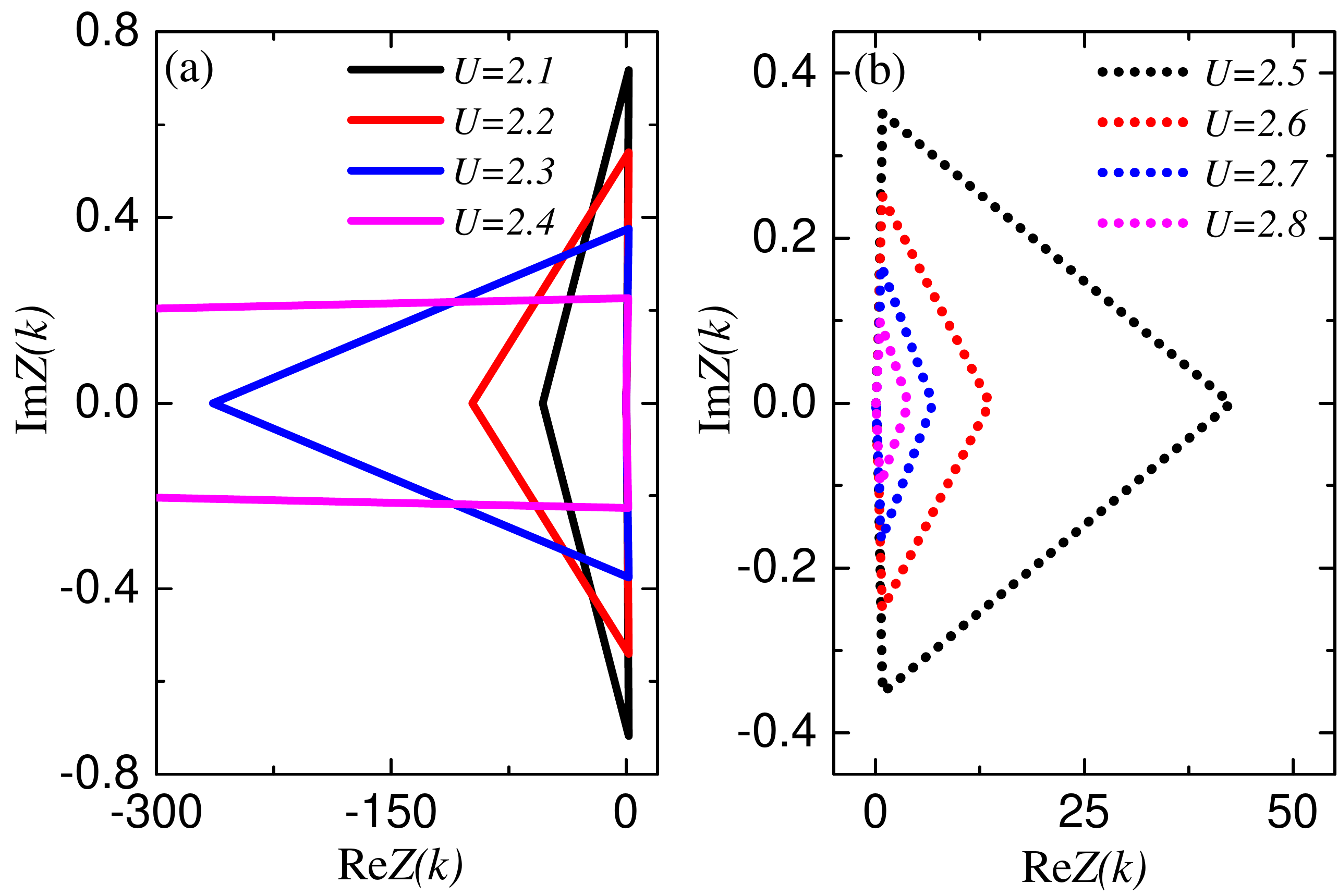}
\end{center}
 \caption{(Color online) The variation of topological invariant from the TSC phase to the CDW phase for the closed interacting Kitaev chain.
          The horizontal (vertical) axis denotes the real (imaginary) part of $Z(k)$ of Eq.~\eqref{Eq_N1_2}.
          The purple loop in the left panel denotes $U=2.4$, which is not plotted completely and the minimal real part of this loop reaches to less than $-3000$.
          The origin is inside (outside) the loci of the left (right) panel.
          The parameters of model are chosen as $\Delta$=0.4, $L$=32.
         }
 \label{Fig_TN_2}
\end{figure}
Fig.~\ref{Fig_TN_2} shows that $Z(k=\pi)$ approach infinity along the negative or positive direction of the real axis if $U$ varies from left side or right side towards the phase boundary at $U_c\approx2.5$. Therefore, there is a pole of eigenvalues of the zero-frequency dimerized Green's functions at $U_c\approx2.5$, which results in the variation of the topological invariant between the topologically trivial CDW phase ($N_1=0$) and the nontrivial TSC phase ($N_1=1$). Therefore, with the help of the generalized formalism, the topological invariant of the CDW phase can be correctly computed.

It is worthwhile to mention that the topological invariant of the interacting Kitaev chain can also be measured by the fermion parity of the ground state.
We find that the ground states of the CAT and CDW phase both reside in the even sector of the Hilbert space while that of the TSC phase lives in the odd sector.
It seems that the topological invariant expressed in terms of the zero-frequency Green's functions is redundant for the interacting model, whose role may be replaced by the topological invariant defined by the fermion parity of the ground state. This motivate us to compare these two ways of computing topological invariants by investigating the interacting dimerized Kitaev chain.

\begin{figure}[hbp]
\begin{center}
  \includegraphics[width=8cm, angle=0]{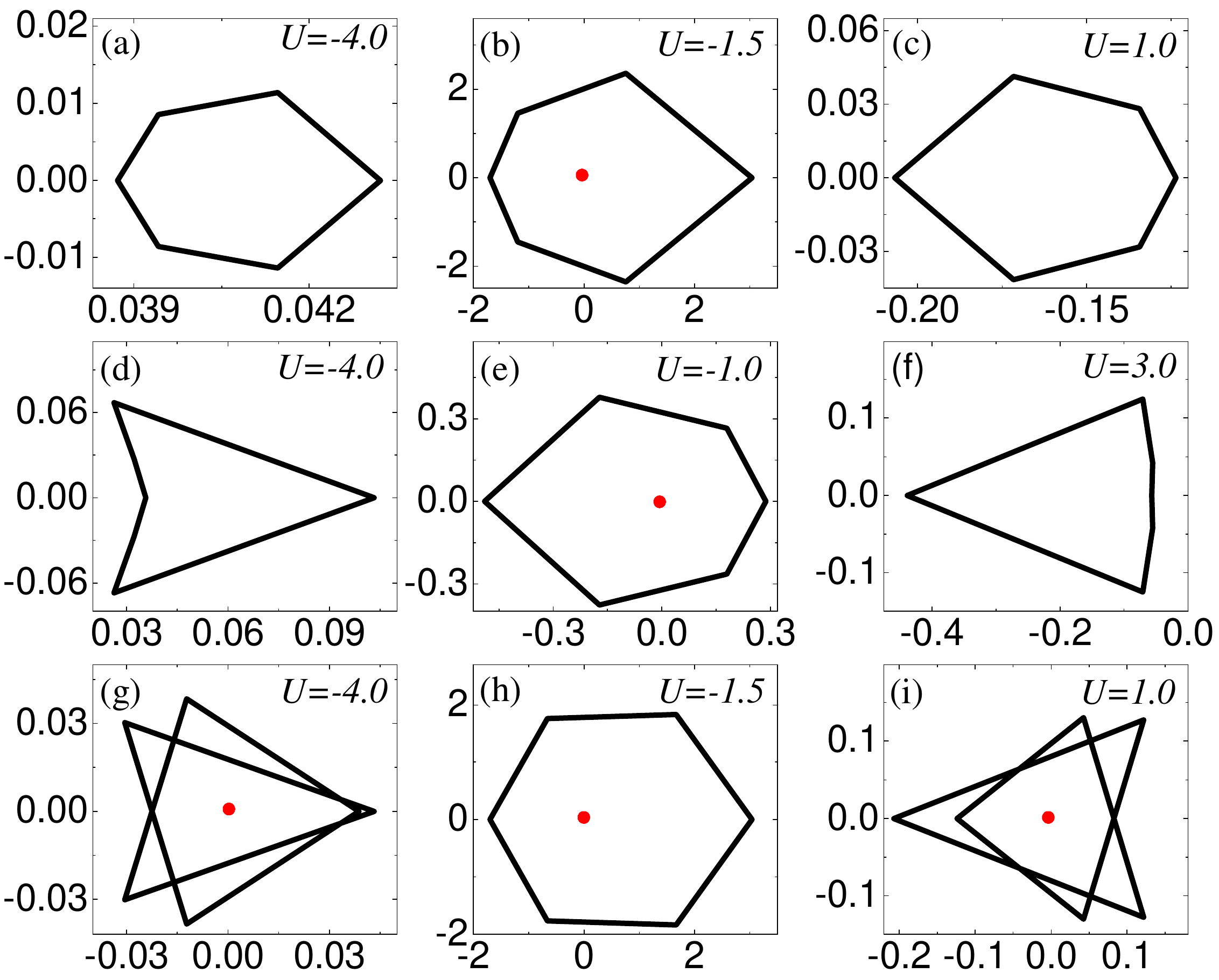}
\end{center}
 \caption{(Color online) The locus of $Z(k)$ for the closed interacting dimerized Kitaev chain for different $U$ in three sets of parameters:
          $(\Delta, \eta) = (0.2,  0.7)$ for (a), (b), (c); $(0.7,  0.2)$ for (d), (e), (f); $(0.2, -0.7)$ for (g), (h), (i).
          The horizontal (vertical) axis of all denote the real (imaginary) part of $Z(k)$.
          The origins are denoted by the red dots.
         }
 \label{Fig_TN_3}
\end{figure}
The interacting dimerized Kitaev chain is a hybrid 1D model which combines the interacting Kitaev chain with the Su-Schrieffer-Heeger (SSH) model. The Hamiltonian is written as,
\begin{align}\label{Eq_Ham_SSH-K-U}
     H^{\prime}   = & - \mu    \sum_{j} ( c_{A,j}^{\dagger}c_{A,j} + c_{B,j}^{\dagger}c_{B,j} )  \nonumber\\
                            & - t      \sum_{j} [(1+\eta) c_{B,j}^{\dagger}c_{A,j} + (1-\eta) c_{A,j+1}^{\dagger}c_{B,j} +h.c. ]  \nonumber\\
                            & - \Delta \sum_{j} [(1+\eta) c_{B,j}^{\dagger}c_{A,j}^{\dagger} + (1-\eta) c_{A,j+1}^{\dagger}c_{B,j}^{\dagger} +h.c. ]  \nonumber\\
                            & + U      \sum_{j} [(1+\eta) (n_{B,j}-\frac{1}{2}) (n_{A,j}-\frac{1}{2}) \nonumber\\
                            & ~~~~~~~~         + (1-\eta) (n_{A,j+1}-\frac{1}{2}) (n_{B,j}-\frac{1}{2}) ]
\end{align}
where $t$, $\Delta$, $\mu$, and $U$ have the same definitions as Eq.~\eqref{Eq_Ham_Kitaev-U}.
$A$ and $B$ denote the sublattice sites in one unit cell, and $\eta$ describes the differences between the intercell and intracell parameters.
The noninteracting case has been investigated by R. Wakatsuki et al.~\cite{Ref_Nagaosa_0} When $\mu=0$, three phases are derived corresponding to the chiral operator $\hat{\mathcal{S}}_1=\tau_0 \otimes \sigma_z$:
(i) SSH-like trivial phase ($N_1=0$) for $t|\eta|>|\Delta|$, $\eta>0$;
(ii) Kitaev-like topological phase ($N_1=1$) for $t|\eta|<|\Delta|$;
(iii) SSH-like topological phase ($N_1=2$) for $t|\eta|>|\Delta|$, $\eta<0$.

\begin{table}[htbp]
 \caption{The topological invariant $N_1$ defined by the Green's functions and the parity of the ground state
          for the closed interacting dimerized Kitaev chain calculated by the exact diagonalization method.
          Three sets of parameters are chosen as:
          $(\Delta, \eta)$ $=$ $(0.2,0.7)$, $(0.7,0.2)$, $(0.2,-0.7)$, and $L=12$.
         }
\newcommand{\tabincell}[2]{\begin{tabular}{@{}#1@{}}#2\end{tabular}}
\begin{tabular}{| c | c | c | c |}
\hline
 $(\Delta,\eta)$ & $U$ & $N_1$ & \tabincell{c}{Parity of the \\ ground state} \tabularnewline
\hline
\multirow{3}{*}{$(0.2,0.7)$} & $U<-2.0$ & $0$ & Even\tabularnewline
\cline{2-4}
 & $-2.0<U<-1.2$ & $1$ & Odd\tabularnewline
\cline{2-4}
 & $U>-1.2$ & $0$ & Even\tabularnewline
\hline
\multirow{3}{*}{$(0.7,0.2)$} & $U<-2.7$ & $0$ & Even \tabularnewline
\cline{2-4}
 & $-2.7<U<+1.7$ & $1$ & Odd\tabularnewline
\cline{2-4}
 & $U>+1.7$ & $0$ & Even\tabularnewline
\hline
\multirow{3}{*}{$(0.2,-0.7)$} & $U<-2.0$ & $2$ & Even\tabularnewline
\cline{2-4}
 & $-2.0<U<-1.2$ & $1$ & Odd\tabularnewline
\cline{2-4}
 & $U>-1.2$ & $2$ & Even\tabularnewline
\hline
\end{tabular}
\label{Table_N1_00}
\end{table}
For the case $\mu=0$ and $U \neq 0$, we calculate the topological invariant defined by the Green's functions (as that of the foregoing CDW phase except that the chiral operator
$\hat{\mathcal{S}}_1$ is used according to Ref.~\cite{Ref_Nagaosa_0}) and the fermion parity of the ground state by the exact diagonalization (ED) method.
Three sets of parameters are chosen: $(\Delta, \eta) = (0.2, 0.7)$, $(0.7, 0.2)$ and $(0.2, -0.7)$,
with the corresponding $N_1$ for $U=0$ are equal to $0$, $1$, $2$ respectively.
Fig.~\ref{Fig_TN_3} shows some typical loci of $Z(k)$, from which we find that the topological invariants expressed in terms of the Green's function are well defined for all the three sets of parameters.
Furthermore, it can be seen from Table~\ref{Table_N1_00} that since ground states of the trivial phase ($N_1=0$) and the topological phase ($N_1=2$) both reside in the even sector of
the Hilbert space, the fermion parity itself can not while the Green's function formalism can distinguish these topologically distinct phases of the interacting dimerized Kitaev chain.

In summary,
we have shown that the $2\times2$ Green function formalism  works well in calculating the topological invariants for the TSC and CAT phases of the interacting Kitaev chain.
However, it has to be generalized to the $4\times 4$ form when the topological invariant of the CDW phase is concerned.
The variation of topological invariant of the interacting Kitaev chain can be attributed to the poles of eigenvalues of the zero-frequency Green's functions, which is the same as the noninteracting Kitaev chain.
The topological invariant of the interacting dimerized Kitaev chain is also calculated, which shows that the Green's function formalism can give more detailed distinction of different topological phases than the fermion parity.

\section*{Acknowledgements}
This work was supported by the National Natural Science Foundation of China under Grant No 11274379, and the Research Funds of Renmin University of China under Grant No 14XNLQ07.

\end{document}